\documentclass[preprint,amssymb,amsmath,amsfonts,russian]{revtex4-1}

\usepackage{graphicx}
\usepackage[russian]{babel}

\begin{document}

 \title{О скорости ударной ионизации в прямозонных полупроводниках}

 \date{21.02.2017}

 \author{А.~Н.~Афанасьев}
 \email{afanasiev.an@mail.ru}
 \author{А.~А.~Грешнов}
 \author{Г.~Г.~Зегря}
 \affiliation{Физико-технический институт им. А.Ф. Иоффе РАН, Политехническая ул., 26, Санкт-Петербург, 194021, Россия}

\begin{abstract}
В рамках 14-зонной ${\bf k}\cdot{\bf p}$ модели исследована интенсивность процессов ударной ионизации в прямозонных полупроводниках и получены аналитические выражения для темпа ударной ионизации. Показано, что вблизи энергетического порога скорость процесса определяется суммой изотропного вклада, кубического по отстройке от порога, и сильно анизотропного квадратичного, возникающего лишь в меру взаимодействия с далекими зонами. Сопоставление этих вкладов в условиях усреднения по невырожденному изотропному распределению неравновесных электронов с некоторой эффективной температурой $T^*$ показывает, что именно кубический, а не традиционно используемый квадратичный вклад доминирует для прямозонных полупроводников с $E_g<1-1.5\,\textrm{eV}$ вплоть до $T^* = 300\,\textrm{K}$, и это должно учитываться при расчетах приборных характеристик устройств, использующих эффект лавинного умножения носителей.
\end{abstract}

\maketitle

\section*{Введение}
\begin{figure}[t!]
\centering
 \includegraphics[width=60mm]{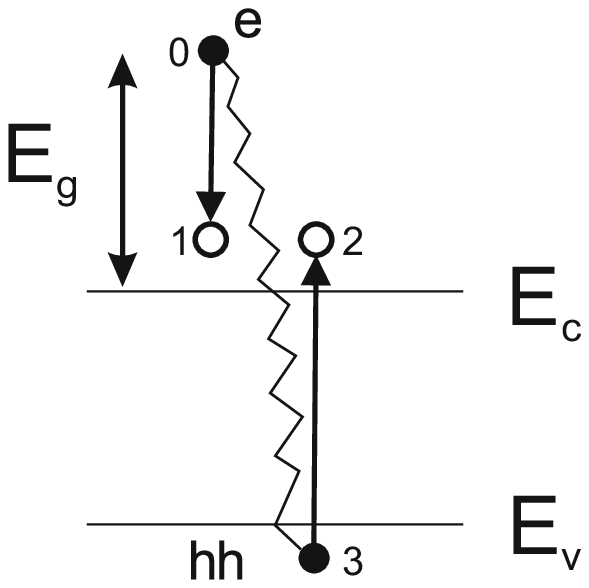}
 \caption{Схема элементарного акта ударной ионизации}
 \label{Fig1}
\end{figure}

Явление ударной ионизации в полупроводниках, состоящее в рождении электрон-дырочной пары в результате кулоновского взаимодействия высокоэнергетичного электрона проводимости с электронами валентной зоны (см. Рис.~\ref{Fig1}), лежит в основе функционирования большого семейства устройств современной электроники, таких как лавинно-пролетные диоды (IMPATT), лавинные фотоприемники (APD)~\cite{Sze}, а также транзисторы с полевым контролем ударной ионизации (I-MOS)~\cite{Gopalakrishan}, в которых экспериментально достигнута крутизна подпороговой части ВАХ на уровне 5\,mV/dec при T=400\,K, что позволяет в разы увеличить скорость переключения. С точки зрения эффективности применения ударной ионизации в качестве физического принципа работы приборов наиболее удачным является использование в качестве областей лавинного умножения рассматриваемых в данной работе прямозонных полупроводников с относительно небольшой шириной запрещенной зоны $E_g\lesssim1\textrm{ eV}$, в которых энергетический порог лишь немного отличается от величины щели: $E_{th}\approx E_g(1+2\mu)$~\cite{APY}, где $\mu=m_e/m_{hh}\ll 1$. Одним из существенных факторов, затрудняющих развитие данных устройств, является недостаточная разработанность теоретического описания происходящих в них процессов ударной ионизации. Вследствие отсутствия последовательно полученных аналитических выражений для полного темпа ударной ионизации, большинство ''инженерных'' расчетов в настоящее время выполняются на основе упрощенных формул со свободно подгоняемыми параметрами, что не может не отражаться на адекватности получаемых результатов. В литературе существует неопределенность не только относительно величины префактора $C$, но и степени $n$ в соотношении
\begin{equation}
 {\mathcal W}(E)=C(E-E_{th})^n,
 \label{eq0}
\end{equation}
описывающем зависимость темпа процесса от энергии налетающего электрона вблизи порога ударной ионизации $E_{th}$. Хотя в целом наиболее популярны представления о квадратичной зависимости $\mathcal{W}(E)$ ($n=2$), предложенной из общих соображений Л.В.~Келдышем еще в 1959~г.~\cite{Keldysh}, авторы работ~\cite{Choo},\cite{Harrison} пришли к выводу, что результаты численных расчетов $\mathcal{W}(E)$ для тройного соединения In$_{0.53}$Ga$_{0.47}$As наилучшим образом описываются с помощью таких степеней, как $n=2.5$~\cite{Choo}, $n=4.3$~\cite{Choo} и $n=5.6$~\cite{Harrison}. Кроме того, как отмечали Гельмонт и соавторы в работе~\cite{Gelmont}, для наиболее узкозонных полупроводников с прямой запрещенной зоной, в частности, рассмотренного ими соединения $\rm{Cd}_{0.2}\rm{Hg}_{0.8}\rm{Te}$ с  $E_g\approx0.1\rm{ eV}$, квадратичный вклад в $\mathcal{W}(E)$ представляет лишь академический интерес, а на деле зависимость $\mathcal{W}(E)$ является кубической, и для нее было приведено явное выражение (к сожалению, без подробностей его вывода).

В данной работе представлен микроскопический анализ процессов ударной ионизации в прямозонных полупроводниках, в результате которого была получена обоснованная функциональная зависимость темпа ударной ионизации $\mathcal{W}$ от $E-E_{th}$. Показано, что скорость процесса определяется суммой изотропного вклада, кубического по отстройке от порога, и сильно анизотропного квадратичного, возникающего лишь в меру взаимодействия с далекими зонами. Сопоставление этих вкладов в реалистичных условиях показывает, что доминирование кубического вклада возникает, начиная с $E_g\sim1\,\textrm{eV}$, и это обстоятельство должно учитываться при численном моделировании полупроводниковых приборов, экплуатирующих эффект лавинного умножения носителей.

\section{Теория.}
Для расчета скорости ударной ионизации будем использовать золотое правило Ферми, определяющее темп элементарного процесса ударной ионизации в результате кулоновского взаимодействия горячего электрона зоны проводимости с заполненной валентной зоной:
\begin{equation}
 W_{\{ 0,3\} \rightarrow \{ 1,2 \}} = \frac{2\pi}{\hbar}
 \left| \left\langle \alpha_1\alpha_2
 \left| \hat{V} \right|
 \alpha_{0}\alpha_{3} \right \rangle \right|^2
 \delta(\Delta \mathcal{E}),
 \label{Prob_comm}
\end{equation}
где нижние индексы нумеруют состояния согласно Рис.~\ref{Fig1}, $\hat{V}$ -- оператор кулоновского взаимодействия, $\Delta \mathcal{E}=E_1+E_2-E_3-E_0$ отражает баланс энергии~\cite{footnoteE0}, а $\alpha_i=\{{\bf k}_i,\xi_i \}$ -- набор квантовых чисел: волнового вектора и спинового индекса. Темп ударной ионизации для заданного состояния налетающего электрона 0 (фигурирующий в левой части формулы~(\ref{eq0})) определяется как
\begin{equation}
 \mathcal{W}=\tau^{-1}=\sum\limits_{\alpha_1,\alpha_2,\alpha_3}W_{\{ 0,3\} \rightarrow \{ 1,2 \}}.
\end{equation}
Поскольку электроны, способные к ударной ионизации, описываются экспоненциальным хвостом функции распределения~\cite{Dmitriev}, затухающим на характерном масштабе порядка нескольких десятков meV, в практическом отношении для описания темпа ударной ионизации достаточно иметь выражение для $\mathcal{W}(E)$ лишь вблизи энергетического порога, где оно складывается из конкурирующих квадратичного и кубического вкладов,
\begin{equation}
\label{ratecommon}
 \mathcal{W} = \mathcal{W}_2 + \mathcal{W}_3 = A(E-E_{th})^2+B(E-E_{th})^3,
\end{equation}
причем их сопоставимость, как будет показано ниже, обуславливается относительной малостью коэффициента $A$ в меру отношения ширины запрещенной зоны $E_g$ к расстоянию до далеких зон. Заметим также, что поскольку разница в пороговых энергиях для процессов ударной ионизации с участием тяжелых и легких дырок близка к ширине запрещенной зоны, вкладом последних можно пренебречь ввиду экспоненциальной малости по параметру $E_g/\mathcal{T}$, где $\mathcal{T}$ -- эффективная температура неравновесной функции распределения горячих электронов.

\begin{figure}[t!]
\centering
\includegraphics[width=60mm]{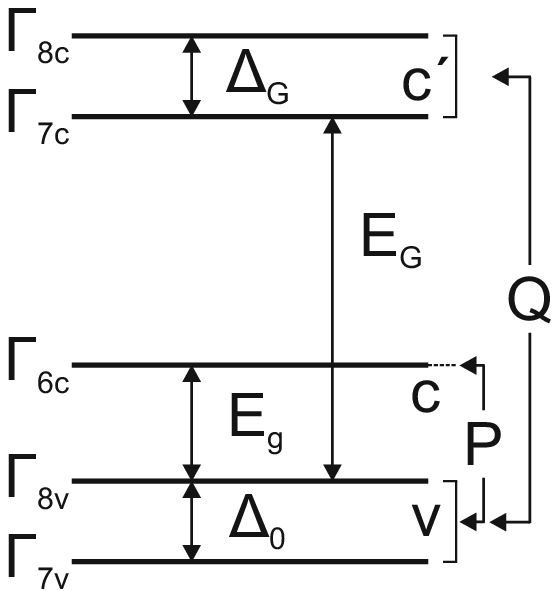}
\caption{Графическое представление 14-зонной ${\bf k}\cdot{\bf p}$ модели} \label{Fig2}
\end{figure}

Для последующего анализа удобно воспользоваться представлением кулоновского взаимодействия в виде интеграла Фурье~\cite{Zegrya}, а волновые векторы участвующих в процессе частиц представить в виде суммы пороговых значений ${\bf k}_i^{th}$ и малых отстроек ${\bf q}_i={\bf k}_i-{\bf k}_i^{th}\ll{\bf k}_i^{th}$. С учетом малости величин $q_i$ выражение для $\mathcal{W}(E)$ преобразуется к виду
\begin{multline}
 {\mathcal W} = \frac{\pi\hbar F\left(\frac{\Delta_{0}}{Eg}\right)}{12m_e E_g^2} \left(\frac{4\pi e^2}{\varkappa}\right)^2
 \int\frac{d^3q_1d^3q_2}{(2\pi)^6}
 [\mathcal{I}_{cv}({\bf q}_1,{\bf q}_3) + \\
 \mathcal{I}_{cv}({\bf q}_2,{\bf q}_3)]
 \delta\left(q^2_1+q^2_2-\frac{2m_e(E_0-E_{th})}{\hbar^2}\right),
 \label{rate_final}
\end{multline}
где $\mathcal{I}_{cv}({\bf q}_i,{\bf q}_j)=I_{cv}({\bf k}_i({\bf q}_i),{\bf k}_j({\bf q}_j))$, $I_{cv}$ обозначают просуммированные по спиновым переменным квадраты интегралов перекрытия блоховских функций тяжелой дырки и конечного электрона, ${\bf q}_3 = {\bf q}_1+{\bf q}_2-{\bf q}_0$ выражается через переменные интегрирования с помощью закона сохранения импульса, причем $q_0=\left(\frac{\partial E_0}{\partial k_0}\right)_{th}^{-1}(E_0-E_{th})$ и сонаправлен с ${\bf k}_0^{th}$, $\Delta_{0}$ -- величина спин-орбитального расщепления валентной зоны, а
\begin{equation}
 F(x)=\frac{(1+x)^2 (1+x/3)^3}{(1+7x/9+x^2/6) (1+2x/3)^2 (1+x/2)}.
\end{equation}
Из формулы~(\ref{rate_final}) следует, что последовательные показатели степени в разложении $\mathcal{W}(E)$ по $E-E_{th}$ соответствуют разложению $\mathcal{I}_{cv}$ по  ${\bf q}_i$. В частности, квадратичный вклад определяется интегралами перекрытия для пороговой конфигурации участвующих в процессе частиц $\mathcal{I}_{cv}({\bf 0},{\bf 0})=I_{cv}({\bf k}_i^{th},{\bf k}_j^{th})$, в соответствии с которой для изотропного закона дисперсии волновые векторы в начальных и конечных состояниях должны быть коллинеарны~\cite{Anderson}, а импульс налетающего электрона в пределе $\mu=m_e/m_{hh}\ll1$ отдается преимущественно дырке~\cite{APY}. Соответственно, вклады с более высокой степенью (в том числе и кубический) определяются неколлинеарным взаиморасположением волновых векторов. Поэтому для определения коэффициентов $A$ и $B$ в формуле~(\ref{ratecommon}) достаточно выполнить расчет вкладов в $\mathcal{I}_{cv}$, происходящих из разложения по отстройкам волновых векторов от пороговых (${\bf q}_i$). При этом мы будем использовать 14-зонную ${\bf k}\cdot{\bf p}$ модель~\cite{Winkler}, явно учитывающую шестикратно вырожденную вторую зону проводимости, которая расположена на несколько eV выше $E_c$ (см. Рис.~\ref{Fig2}). Если же ограничиться учетом лишь ${\bf k}\cdot{\bf p}$ взаимодействия между зоной проводимости (c) и валентной зоной (v), полученный интеграл перекрытия будет соответствовать нулевой величине квадратичного вклада, т.к. блоховские функции тяжелой дырки и электрона ортогональны при коллинеарных волновых векторах в рамках восьмизонной модели Кейна, не включающей слагаемые латтинжеровского типа~\cite{Kane},~\cite{Greshnov}. При этом выражение для коэффициента $B$ имеет вид
\begin{equation}
 B = \frac{\omega_B^*}{18E_g^3} \frac{E_g+\Delta_{0}}{E_g+\frac{2}{3}\Delta_{0}}
 F\left(\frac{\Delta_{0}}{E_g}\right),
 \label{B}
\end{equation}
которое в пределе $\Delta_{0}\gg E_g$ переходит в приведенное в работе~\cite{Gelmont} (здесь $\omega_B^*=m^*e^4/2\varkappa^2\hbar^3$ -- боровская частота для электронов зоны проводимости). Заметим, что кубический вклад не зависит от направления волнового вектора налетающего электрона, поскольку он происходит от сферически симметричной части 14-ти зонного ${\bf k}\cdot{\bf p}$ гамильтониана.

\begin{figure}[t!]
 \includegraphics[]{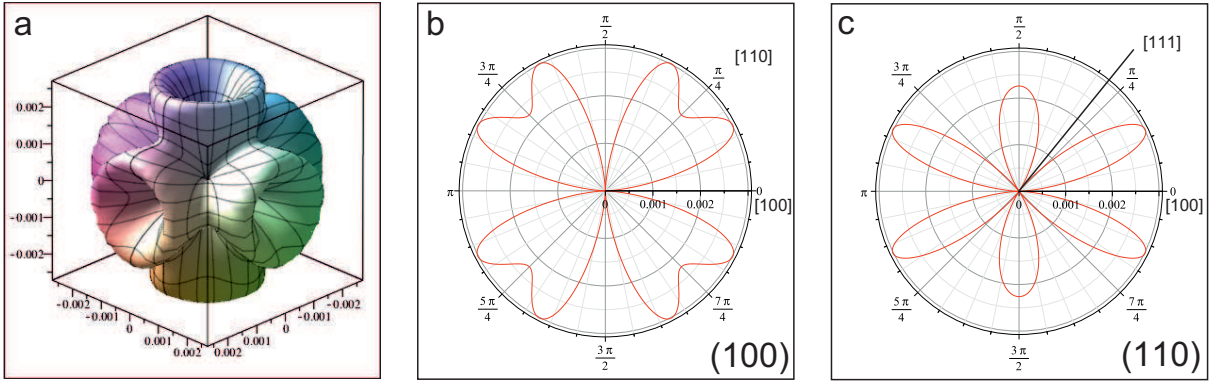}
 \caption{Угловая зависимость квадратичного вклада в темп ударной ионизации $\mathcal{W}_2(\Theta_0,\varphi_0)$ в единицах $\omega_B^*$ для заданной энергии налетающего электрона $(E-E_{th})/E_g=\sqrt{3/2}$. (а) Трехмерный график (b) Сечение плоскостью (100) (c) Сечение плоскостью (110).}
 \label{Fig3}
\end{figure}

Для расчета квадратичного вклада удобно воспользоваться малостью отношения величин $E_g, \Delta_0, \Delta_G$ к расстоянию между валентной зоной (v) и второй зоной проводимости (c$'$), и в результате расчета $\mathcal{I}_{cv}$ по теории возмущений мы приходим к следующему выражению для коэффициента при квадратичном вкладе:
\begin{equation}
 A = \frac43 \frac{\omega_B^*}{E_G^2} \frac{Q^4}{P^4} J_4({\bf u})
 \frac{E_g+\frac12\Delta_0}{E_g+\frac13\Delta_0} F\left(\frac{\Delta_{0}}{E_g}\right),
 \label{A}
\end{equation}
где P и Q -- матричные элементы оператора импульса между блоховскими функциями зон c-v и c$'$-v соответственно (Рис.~\ref{Fig2}), $J_4({\bf u})=I_1({\bf u})-3I_1^2({\bf u})$ -- комбинация кубических инвариантов четвертого порядка $I_1({\bf u}) = u_x^2u_y^2 + u_x^2u_z^2 + u_y^2u_z^2$, ${\bf u}={\bf k}_0/k_0$ -- единичный вектор в направлении движения начального электрона. Данный вклад возникает в меру подмешивания к блоховских функции тяжелой дырки состояний далеких зон, которое в рамках 14-ти зонной модели описывается блоком ${\bf k}\cdot{\bf p}$-взаимодействия валентной зоны со второй зоной проводимости (c$'$), имеющим кубическую симметрию. Вследствие этого квадратичный вклад приобретает нетривиальную угловую зависимость, продемонстрированную на Рис.~\ref{Fig3}. В частности, квадратичный вклад исчезает, если инициирующий акт ударной ионизации  электрон налетает в кристаллографических направлениях [100] или [111]. Также заметим, что коэффициент $A$ содержит в знаменателе вторую степень $E_G$ и является в этом смысле малым по сравнению с коэффициентом $B$, поэтому конкурентность полученных вкладов, доминирование квадратичного или кубического, зависит от конкретного распределения горячих электронов по энергии.

\section{Обсуждение и выводы.}
В практическом отношении важную роль играет интегральная скорость ударной ионизации
\begin{equation}
 \mathcal{R} = \sum\limits_{\alpha_0} \mathcal{W}(E({\bf k}_0)){\tilde f}_{\alpha_0},
 \label{eq_totrate}
\end{equation}
имеющая смысл числа дополнительных электрон-дырочных пар, генерируемых в единицу времени в результате элементарных процессов ударной ионизации, и зависящая от высокоэнергетического хвоста неравновесной добавки к функции распределения ${\tilde f}_{\alpha}=f_{\alpha}-f_0(E({\bf k}))$. Поскольку нечетные гармоники функции распределения (в том числе и главная, генерируемая электрическим полем) согласно формуле~(\ref{eq_totrate}) дают нулевой вклад в величину $\mathcal{R}$, для сопоставления величины двух вкладов в интегральную скорость ударной ионизации мы сравним парциальные вклады $\mathcal{R}_{2,3}$, происходящие от двух слагаемых в формуле (\ref{ratecommon}) при неравновесной добавке к функции распределения вида ${\tilde f}(E)=\mathcal{N} \exp(-E/\mathcal{T})$. Здесь $\mathcal{T}$ обозначает эффективную температуру распределения горячих электронов (совпадающую в случае слабого внешнего воздействия с температурой решетки). В этом случае выражения для вкладов $\mathcal{R}_2$ и $\mathcal{R}_3$ можно записать в виде
\begin{equation}
 {\mathcal R}_i =
 \mathcal{D}(E_{th}) \mathcal{N} \int_{E_{th}}^{\infty} \overline{{\mathcal W}_i(E)} \exp(-E/{\mathcal T}) dE,
 \label{eq_Ri}
\end{equation}
где горизонтальная черта обозначает усреднение по углам, а $\mathcal{D}(E)$ -- плотность состояний. Используя величину среднего $\overline{J_4({\bf u})}=2/35$ и приравнивая получающиеся выражения для $\mathcal{R}_{2,3}$, для эффективной температуры кроссовера между двумя механизмами имеем
\begin{equation}
 {\mathcal T}^*=\frac{\overline{A}}{3B}=\frac{16}{35}\frac{Q^4}{P^4}\frac{E_g^3}{E^2_G}G\left(\frac{\Delta_0}{E_g}\right),
 \label{Teff}
\end{equation}
где $G(x)=\frac{(1+2x/3)(1+x/2)}{(1+x)(1+x/3)}$. Для параметров полупроводника InAs~\cite{Winkler}, обладающего запрещенной зоной около $0.4\,\textrm{eV}$, данное выражение дает ${\mathcal T}^*\approx10\,\textrm{K}$, соответственно при комнатной температуре квадратичный вклад в 50 раз меньше кубического. Как несложно видеть из формулы (\ref{Teff}), доминирование кубического вклада над квадратичным при ${\mathcal T}^*=300\,\textrm{K}$ ожидается для большинства прямозонных полупроводников с $E_g \lesssim 1-1.5\,\textrm{eV}$, при этом с количественной точки зрения полное пренебрежение кубическим вкладом вряд ли оправдано и для более широкозонных материалов.

Необходимо заметить, что хотя предлагаемая аналитическая теория темпа ударной ионизации при анализе квадратичного вклада использует в качестве малого параметра отношение ширины запрещенной зоны $E_g$ к расстоянию от валентной зоны до второй зоны проводимости $E_G$ и ее результаты кажутся применимыми для произвольного отношения $\Delta_0/E_g$, это не совсем так. Дело в том, что при уменьшении спин-орбитального расщепления валентной зоны дисперсионная ветвь спин-орбитально отщепленных дырок $E_{so}({\bf k})$ приближается к ветви тяжелых дырок и в пределе $\Delta_{0}\rightarrow0$ совпадает с ней, образуя вторую ветвь тяжелых дырок. В таких условиях возникает вырождение, препятствующее применимости использованной в работе формы теории возмущений. Поэтому на самом деле формула (\ref{A}) корректна при условии $\Delta_0/E_g \gg E_g/E_G$, что на практике может не выполняться для полупроводников из относительно легких элементов с запрещенной зоной более 1\,eV, например для наиболее популярного прямозонного полупроводника GaAs параметр $\beta=\Delta_0 E_G/E_g^2$ составляет 0.8, а для InP -- 0.25. Однако, как показывают проведенные нами численные расчеты, усредненная по углам величина коэффициента $\overline{A}$ отличается от соответствующей формуле (\ref{A}) не более чем на 10\,\%, хотя угловая зависимость при этом существенно видоизменяется. На основании проведенного расчета можно утверждать, что и для указанных соединений кубический вклад в интегральный темп ударной ионизации сопоставим по величине с квадратичным при эффективной температуре неравновесной функции распределения горячих электронов порядка 300\,K. Также, в ближайшее время мы планируем построить аналитическую теорию, описывающую квадратичный вклад в темп ударной ионизации для случая $\beta\ll1$, трактуя при этом спин-орбитальное взаимодействия в качестве возмущения.

\end{document}